\documentclass[twoside,a4paper,11pt]{article}
\usepackage{times}
\usepackage{elcvia}
\usepackage[english]{babel}
\usepackage{tcolorbox}
\tcbuselibrary{breakable}
\usepackage{amssymb}
\usepackage{graphicx}
\usepackage{wrapfig} 
\usepackage{float}
\usepackage[colorlinks,linkcolor=blue]{hyperref}

\newcommand{\minisection}[1]{\vspace{.06in}\noindent{\textbf{#1}}.}
\input{psfig.sty}

\title{ChipExpert: The Open-Source Integrated-Circuit-Design-Specific Large Language Model 
\footnote[0]{Correspondence to: xning@seu.edu.cn}
\footnote[0]{\noindent Principal Investigator: Ning Xu} 
}

\author{Ning Xu$^{1,2}$ ~ Zhaoyang Zhang$^{1,2}$ ~ Lei Qi$^{1,2}$ ~ Wensuo Wang$^1$ ~ Chao Zhang$^1$ ~ Zihao Ren$^2$\\
Huaiyuan Zhang$^2$ ~ Xin Cheng$^2$ ~ Yanqi Zhang$^2$ ~ Zhichao Liu$^2$ ~ Qingwen Wei$^2$ ~ Shiyang Wu$^{1,2}$\\ 
Lanlan Yang$^1$ ~ Qianfeng Lu$^2$ ~ Yiqun Ma$^2$  ~ Mengyao Zhao$^2$ ~ Junbo Liu$^2$ ~ Yufan Song$^1$\\
Xin Geng$^2$ ~ Jun Yang$^{1,2}$\\ \\
\centerline{\small \em $^1$ National Center of Technology Innovation for EDA, Nanjing, China} \\
\centerline{\small \em $^2$ Southeast University, Nanjing, China} \\ \\
\centerline{\small Code: \href{https://github.com/NCTIE/ChipExpert}{https://github.com/NCTIE/ChipExpert}} \\
\centerline{\small Model: \href{https://huggingface.co/China-NCTIEDA/ChipExpert-8B-Instruct}{https://huggingface.co/China-NCTIEDA/ChipExpert-8B-Instruct}} \\
\centerline{\small Benchmark: \href{https://huggingface.co/datasets/China-NCTIEDA/ChatICD-Bench}{https://huggingface.co/datasets/China-NCTIEDA/ChatICD-Bench}} \\
       }

\begin{document}

\maketitle

\pagestyle{myheadings}

\hrulefill

\begin{abstract}

The field of integrated circuit (IC) design is highly specialized, presenting significant barriers to entry and research and development challenges. Although large language models (LLMs) have achieved remarkable success in various domains, existing LLMs often fail to meet the specific needs of students, engineers, and researchers. Consequently, the potential of LLMs in the IC design domain remains largely unexplored. To address these issues, we introduce \textbf{ChipExpert}, the first open-source, instructional LLM specifically tailored for the IC design field. ChipExpert is trained on one of the current best open-source base model (Llama-3 8B). The entire training process encompasses several key stages, including data preparation, continue pre-training, instruction-guided supervised fine-tuning, preference alignment, and evaluation. In the data preparation stage, we construct multiple high-quality custom datasets through manual selection and data synthesis techniques. In the subsequent two stages, ChipExpert acquires a vast amount of IC design knowledge and learns how to respond to user queries professionally. ChipExpert also undergoes an alignment phase, using Direct Preference Optimization, to achieve a high standard of ethical performance. Finally, to mitigate the hallucinations of ChipExpert, we have developed a Retrieval-Augmented Generation (RAG) system, based on the IC design knowledge base. We also released the first IC design benchmark \textbf{ChipICD-Bench}, to evaluate the capabilities of LLMs across multiple IC design sub-domains. Through comprehensive experiments conducted on this benchmark, ChipExpert demonstrated a high level of expertise in IC design knowledge Question-and-Answer tasks. 


\end{abstract}
\hrulefill

\section{Introdution}

The integrated circuit (IC) field is a cornerstone of modern technology, crucial for advancing various industries including telecommunications, computing, and consumer electronics. The highly specialized nature of IC design presents substantial barriers in research and development. Access to comprehensive and in-depth IC design knowledge is often limited, presenting significant challenges for students striving to acquire foundational knowledge and for experts seeking to remain abreast of the latest advancements. Traditional educational resources and training methods often fail to provide the necessary range and depth of knowledge, which results in a steep learning curve for students and high training costs for engineers. The lack of accessible, high-quality instructional materials slows down overall progress and innovation in the IC design industry.

\begin{figure}
    \centering
    \includegraphics[width=1.0\linewidth]{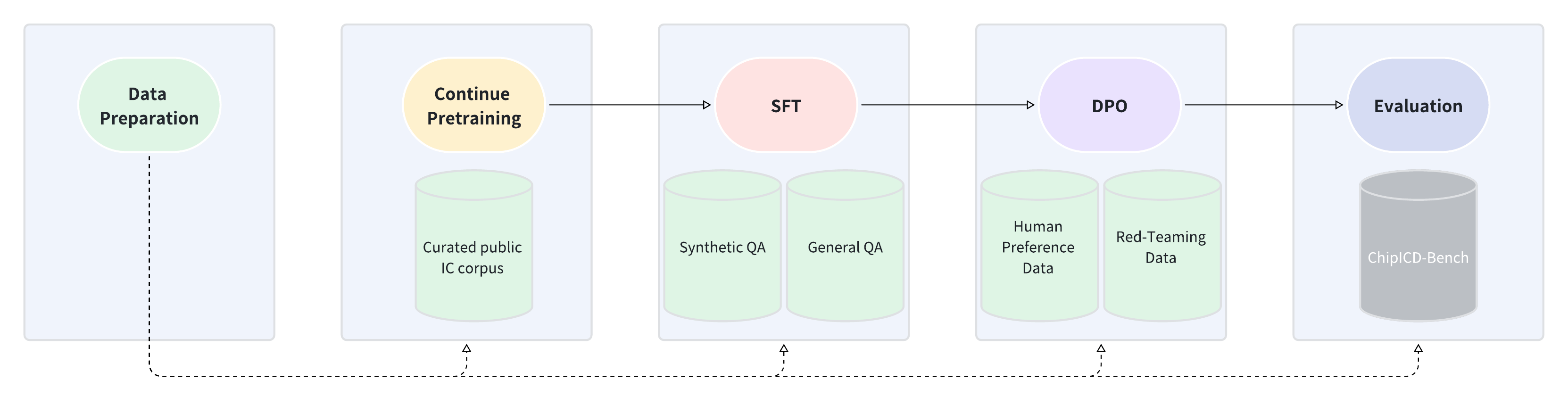}
    \caption{Summary of ChipExpert training stages and data sources.}
    \label{fig:enter-label}
\end{figure}

Large language models (LLMs)\cite{Achiam2023}\cite{Touv2023}\cite{Touvron2023} have recently transformed the landscape of natural language processing (NLP), demonstrating remarkable capabilities across various domains. These models have shown the potential to enhance learning and research efficiency by providing accurate and contextually relevant information. In response to specialized needs, many industry-specific LLMs \cite{Wu2023,Bi2023,Wu2024} have emerged. However, existing LLMs are primarily general-purpose and lack the domain-specific knowledge essential for addressing the intricate needs of IC design experts. This gap highlights the necessity for an LLM specifically tailored to the IC design field.

To bridge this gap, we introduce ChipExpert, the first open-source, instructional LLM dedicated to the Integrated-Circuit-Design industry. Trained on the Llama-3 8B, the model aims to provide teaching assistant services for students in the field of IC to learn fundamental knowledge, engineers to inquire about technical details, and researchers to investigate cutting-edge papers and research topics. The ultimate goal of ChipExpert is to reduce the learning barrier for students and lower the training costs for engineers, thereby fostering innovation and efficiency in the integrated circuit industry.

As depicted in Figure 1,  The training process of ChipExpert is divided into several successive phases:
\begin{itemize}
    \item \textbf{Data Preparation}: According to the Scaling Laws for Neural Language Models\cite{{Kaplan2020}}, data is one of the critical factors in ensuring the performance of LLMs. We place great emphasis on the quality, diversity, and quantity of the data involved in training. We have built multiple custom datasets through manual selection and data synthesis. For continue pre-training, we have an IC-related long text corpus dataset. For SFT, we have a general question-answer pair dataset and a synthesized domain question-answer pair dataset. We then mix the two datasets in a specific ratio to form a complete question-answer pair dataset. For alignment, in addition to the limited amount of high-quality, publicly available paired preference datasets, we also have a Red-Teaming dataset to ensure that the model responds as expected. Besides the training dataset, we carefully constructed a ChipICD-Bench to evaluate the model's performance. 
    \item \textbf{Continue Pretraining}: Continue pretraining\cite{{Gururangan2020}} in LLMs is a technique that involves further training an already pre-trained model on additional data or tasks to improve its performance and adapt it to specific domains or applications. Just like the pretraining of foundation model, continue pretraining is also an autoregressive training process on long-text corpora. Through continue pretraining, ChipExpert has learned a large amount of foundational knowledge and cutting-edge information related to IC design, gaining a deep understanding of the chip industry. Next, it needs to learn how to express its professional knowledge when users inquire about it.
    \item \textbf{Supervised Fine-tuning}: Unlike continue pretraining process, which is an unsupervised training process on long-text corpora, supervised fine-tuning (SFT), also known as domain-specific assistant adaptation or instruct tuning, is a supervised learning process on instruction-response samples. We train ChipExpert on the IC design related question-answer pair dataset, and as the training loss decreases, the model gradually learns how to utilize its domain knowledge to respond to users' domain-specific questions.
    \item \textbf{Direct Preference Optimization}: In the alignment phase, ChipExpert undergoes refinement through the incorporation of human preferences to mitigate potential risks and adverse impacts on individual users and society at large. Common possible ways of doing so include Reinforcement Learning from Human Feedback (RLHF)\cite{Kaufmann2023} and Direct Preference Optimization (DPO)\cite{Rafailov2024}. Due to the tremendous computational resources required for RLHF and the significant instability inherent in the reinforcement learning process, We opt for DPO to accomplish the alignment process of ChipExpert. In our alignment training, we employ a two-phase approach. The initial phase involves utilizing a curated set of high-quality, publicly accessible paired preference data. Subsequently, in the second phase, we undertake a comprehensive red teaming initiative to identify and isolate potentially harmful, unsafe, or illegal model outputs. The insights gleaned from this process are then leveraged to generate targeted alignment data, which is used to mitigate these undesirable responses and refine the model's behavior.
    \item \textbf{Evaluation}: The evaluation process is an integral component of the iterative model training cycle. Only by constructing high-quality benchmarks and conducting rigorous testing can we ascertain the real progress of the training process, gauge the model's mastery of IC design knowledge, and determine whether its ability to address users' technical queries has been enhanced. Relying solely on the training loss curve to assess a model's learning progress can be misleading. A significant decrease in training loss may merely indicate that the model has overfit to the training data, rather than achieving a genuine improvement in its problem-solving capabilities. A comprehensive evaluation is essential for ensuring the model's continuous improvement and its capacity to provide accurate and relevant responses to specialized questions in the field of IC design.
\end{itemize}








\section{Data}

This section introduces the datasets utilized throughout the various training stages of ChipExpert, along with the corresponding data processing pipeline. We focus on three critical dimensions of datasets: quality, diversity, and quantity. These factors are carefully considered throughout our training process. We detail the pipeline of the data processing for continue pretraining in §2.1 and introduce the thorough synthetic data generation step for supervised finetuning in §2.2. In §2.3, we explain the preference alignment datasets generation guided by red teaming efforts, which is of paramount importance to alignment phase. 

\subsection{Continue Pretraining Data}

To ensure the success of continue pre-training, a substantial amount of long-form text data related to IC design is required. In order to guarantee an adequate quantity of data, we have made extensive efforts to collect publicly available text data within the IC domain. To maintain the quality of the data, we have not only focused on acquiring high-quality documents from within the industry but have also employed a complex data cleansing and processing pipeline. Furthermore, to ensure data diversity, we have sourced our data from a wide range of materials, including books, research papers, and manuals, covering various common sub-domains within the IC field.

The dataset is entirely composed of data from publicly available sources, amounting to a total of 4.7 billion tokens. It encompasses ten key areas: digital circuit design, analog circuit design, radio frequency (RF) circuit design, power device design, system-on-chip (SoC), electronic design automation (EDA), RF antenna design, compute-in-memory (CIM), semiconductor fabrication, and neural networks. These areas are covered through five primary sources:

\begin{itemize}
\item \textbf{Textbooks}: Comprehensive educational materials covering fundamental and advanced IC design concepts.
\item \textbf{Research papers}: Journal and conference papers providing the latest research findings and technological advancements in the field.
\item \textbf{Circuit code}: Circuit code repositories and examples, including Verilog and other hardware description languages, essential for understanding practical circuit design and implementation.
\item \textbf{Engineering manuals}: Detailed handbooks and guides used by professionals in the IC design industry, offering practical insights and methodologies.
\item \textbf{Webpages}: Content from webpages related to IC design, providing additional context and diverse perspectives on various topics. Over 20,000 pages related to circuits and electronics were extracted from Wikipedia using keyword mining techniques.
\end{itemize}

The composition and ratio of the pre-training data play a significant role in the model's performance. We divided all pre-training data into domain knowledge, code, and Wiki. Based on our experiments, we found that repeating the domain knowledge four times and mixing it with other sources achieves the best results. The final ratio of blended pre-training data is shown in Table \ref{tab:training_costs}.

\begin{table}[ht]
    \small
    \centering
    \begin{tabular}{l|l|c|c|c}
        \hline
        Data Type & Source & Original Tokens (B) & Training Tokens (B) & Ratio \\
        \hline
        Domain knowledge & Textbooks, Papers, etc. & 2.8 & 11.2 & 0.85 \\
        Code & Verilog code, etc. & 0.6 & 0.6 & 0.05 \\
        Wiki & Wikipedia &1.3 & 1.3  & 0.10 \\
        \hline
    \end{tabular}
    \caption{Tokens and ratio for different continue pretraining data types}
    \label{tab:training_costs}
\end{table}

The rest of this subsection describes the different steps of processing performed on the original domain corpora data, as illustrated in Figure 2. We utilize the open-source tool Marker\footnote{https://github.com/VikParuchuri/marker} to convert the content of literature files into plain text (i.e. PDF Parsing). During the data cleaning process, we apply regular expressions to filter out figures, tables, headers, footers, page numbers, URLs, and references. Additionally, any extra spaces, line breaks, and other non-text characters are removed. Note that special characters, emoticons, and garbled characters are also replaced or eliminated during this process. Followed by the deduplication process, during which we employ hash-based methods\cite{Rao2016} to de-duplicate the data. Deduplication helps reduce the risk of over-fitting during continue pretraining and enhances LLM's generalization capability. Due to the presence of code snippets within the dataset, and the critical importance of code formatting, we have meticulously standardized and formatted the different types of code according to their respective syntax rules.

\begin{figure}
    \centering
    \includegraphics[width=0.8\linewidth]{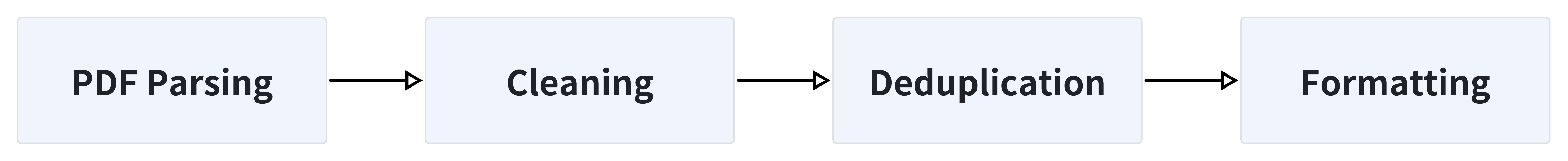}
    \caption{Pipeline of the data processing.}
    \label{fig:enter-label}
\end{figure}

By leveraging this diverse and comprehensive dataset, ChipExpert establishes a robust foundational understanding of the IC design field, ensuring the model is well-equipped to handle a wide range of IC design tasks with high accuracy and relevance.

\subsection{Supervised Finetuning Data}

\begin{figure}
    \centering
    \includegraphics[width=0.59\linewidth]{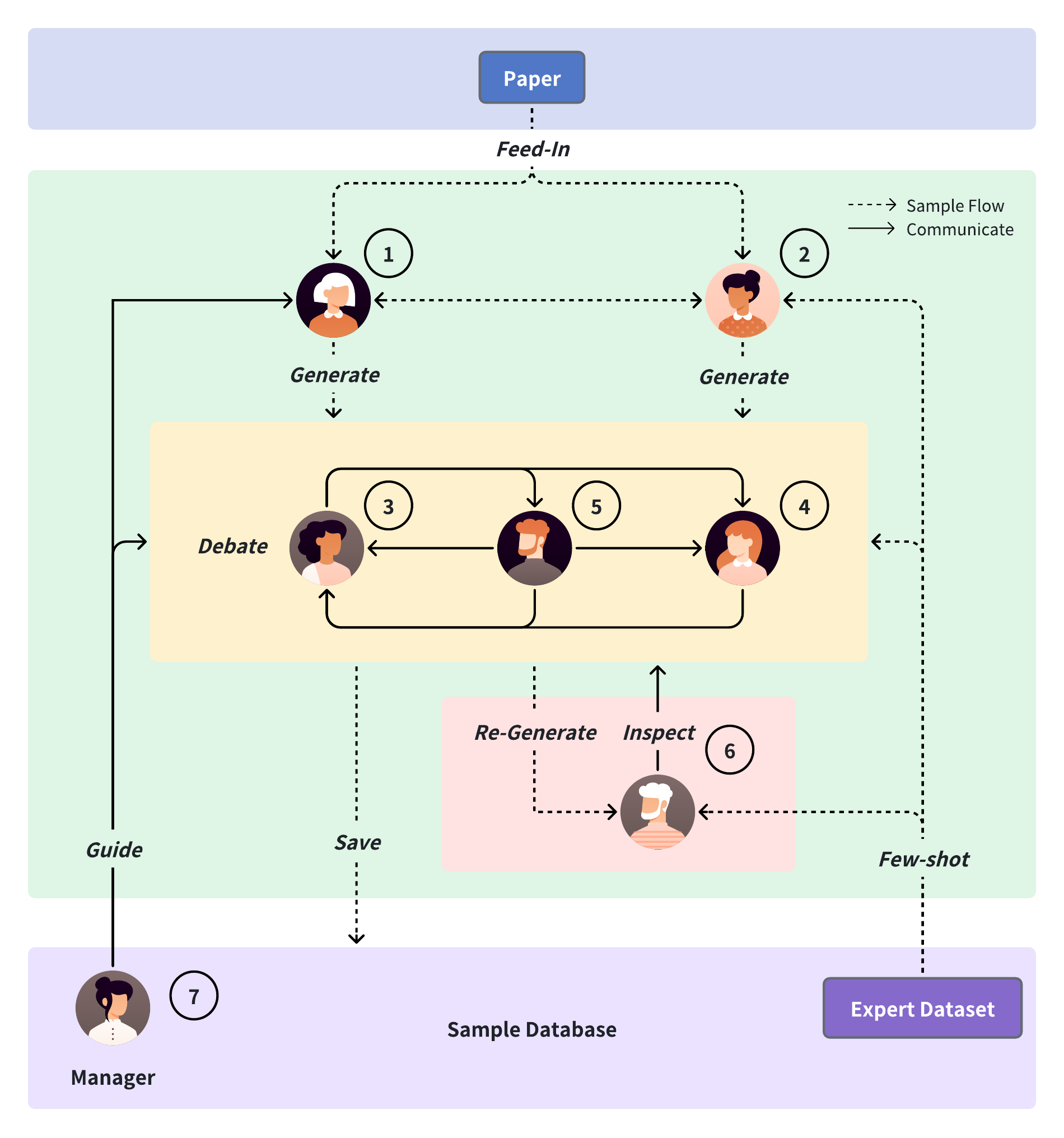}
    \caption{ChipInstruct - Automated Question-Answer Pair Generation Framework}
    \label{fig:enter-label}
\end{figure}

To enable ChipExpert to professionally respond to user queries related to IC design, we require a substantial amount of high-quality question-answer pair data. Our supervised finetuning dataset comprises over 70,000 question-answer pairs. This dataset includes: 
\begin{itemize}
    \item QA pairs constructed by experts: manually designed by domain experts to ensure accuracy and comprehensiveness.
    \item QA pairs constructed from IC forum data: extracted and cleaned from IC design forums to provide real-world scenarios and solutions.
    \item QA pairs adapted from single-choice questions: adapt single-choice questions into QA pairs with complex reasoning processes in the answers, by using GPT-4 and COT prompt.
    \item QA pairs automatically generated with \textbf{ChipInstruct}.
\end{itemize}

ChipInstruct, as illustrated in Figure 3, is an innovative framework specially designed to automate the generation of QA pairs by extracting information from books and papers through multi-agent collaboration. The agent here refers to GPT-4 with complex prompt engineering. Different roles within the framework collaborate to ensure that the dataset is comprehensive, high-quality, and diverse. The rest of this subsection provides a detailed description of the workflow for this collaborative framework:
\begin{enumerate}
    \item A textbook or research paper is first fed to two QA generation agents, Agent$^1$ and Agent$^2$. Agent$^1$ extracts QA pairs heuristically based on the paper, while Agent$^2$ receives the questions generated by Agent$^1$ and performs similarity retrieval within the paper to find the most relevant paragraphs to answer the questions. Agent$^2$ then generates answers and forms complete QA pairs.
    \item The QA pairs generated by the two QA generation agents are then respectively handed over to the two debating agents, Agent$^3$ and Agent$^4$, in the debate module. Agent$^5$ initially acts as a referee. The debating parties argue for their respective answers to the same question, defending their answers by listing the advantages of their own answers, pointing out the deficiencies of the opposing answers, and citing relevant paragraphs from the original textbook or paper as evidence. Agent$^5$ evaluates the debating records of both parties and compares them with the original paper to determine which answer is superior and saves it to the sample database. This triggers Agent$^5$'s second task, Re-Generate. 
    \item As the referee Agent$^5$ witnessing the entire debating process, Agent$^5$ has developed its own understanding of the strengths and weaknesses of the QA pairs and how to generate better ones. Agent$^5$ is then required to enrich and expand the previous QA pairs (in terms of length, vocabulary diversity, splitting into two consecutive questions, etc.) while switching its role to a debater. One of Agent$^3$ and Agent$^4$ becomes the referee for the next round of debate... Eventually, all three agents in the debate module have played the roles of both debater and referee, debating with different roles, and gradually developing the ability to discern the quality of QA pairs and generate better ones.
    \item Due to the uncontrollable nature of LLMs in long-chain communication, we introduce an inspect module. Agent$^6$ in this module determines the number of iterations in the debating module and intervenes multiple times under the guidance of expert data pairs to inspect the newly generated QA pairs in Re-Generate, preventing the debate module from going astray and falling into a vicious cycle.
    \item The final qualified QA pairs are stored in the sample database. Agent$^7$ summarizes, organizes, and analyzes the generated QA pairs for the current textbook or paper, considering whether the diversity is sufficient (balance between concise and detailed, variety of question and answer forms, etc.) and whether the remaining information in the textbook or paper can support the generation of more QA pairs (especially for single-page papers, from which the number of QA pairs that can be extracted is highly limited). After thorough consideration, Agent$^7$ guides the next round of generation for Agent$^1$ and Agent$^2$ by modificating their prompt, and also provides some guidance for the Re-Generate process in the debate module.
\end{enumerate}

When comprehending this framework, it is essential to consider the following key points:
\begin{itemize}
    \item Expert-curated exemplar QA pairs are used throughout the entire process. Each agent makes use of them as a demonstration. As high-quality QA pairs accumulate in the sample database, they will also be sampled and used as a demonstration.
    \item Agent$^7$, acting as the Manager responsible for overseeing the entire sample database, also determines the type of QA pairs (basic conceptual, logical, divergent, etc.) that need to be generated in the current round.
    \item In the multi-agent framework diagram, the dotted arrows represent the generation, sampling, and transfer of QA pair samples. The solid arrows primarily illustrate the communication and message passing process between agents.
\end{itemize}

Inspired by the work of \cite{Dong2023}, we combine our domain-specific QA pair dataset with a small set of high-quality general instruction-tuning datasets using a mixing ratio of 8:1 for fine-tuning the ChipExpert, in order to mitigate the catastrophic forgetting problem\cite{Lin2023}. To accommodate training requirements, we convert the raw datasets into a unified structured format. For single-turn QA, we adopt the Alpaca format, while for multi-turn QA, we employ the ShareGPT format.

\subsection{Preference Alignment Data}

Our alignment training is a two-step process. We collect publicly available paired preference data for the first step. In the second step, we conduct a red teaming effort to identify harmful, unsafe, or illegal responses of ChipExpert, and use these insights to produce alignment data to mitigate them. To generate a dataset of adversarial prompts, we compile and curate entries from Anthropic Harmless \cite{Ganguli2022}, then divide them into train and test splits. We use the ChipExpert to answer the adversarial questions of the training set, and Llama Guard 2\cite{Inan2023} to classify them as either safe or unsafe. We compile the unsafe responses alongside refusals to answers generated with GPT-4, to craft a DPO dataset that covers ChipExpert’s weak points. Given that our human judgments may disagree with the safety/unsafety classifications of Llama Guard 2, We manually review the dataset and remove the defective samples. Finally, We use this dataset for the second step of our alignment training. 

\begin{wrapfigure}{r}[0cm]{0pt}     
    \includegraphics[width=0.4\linewidth]{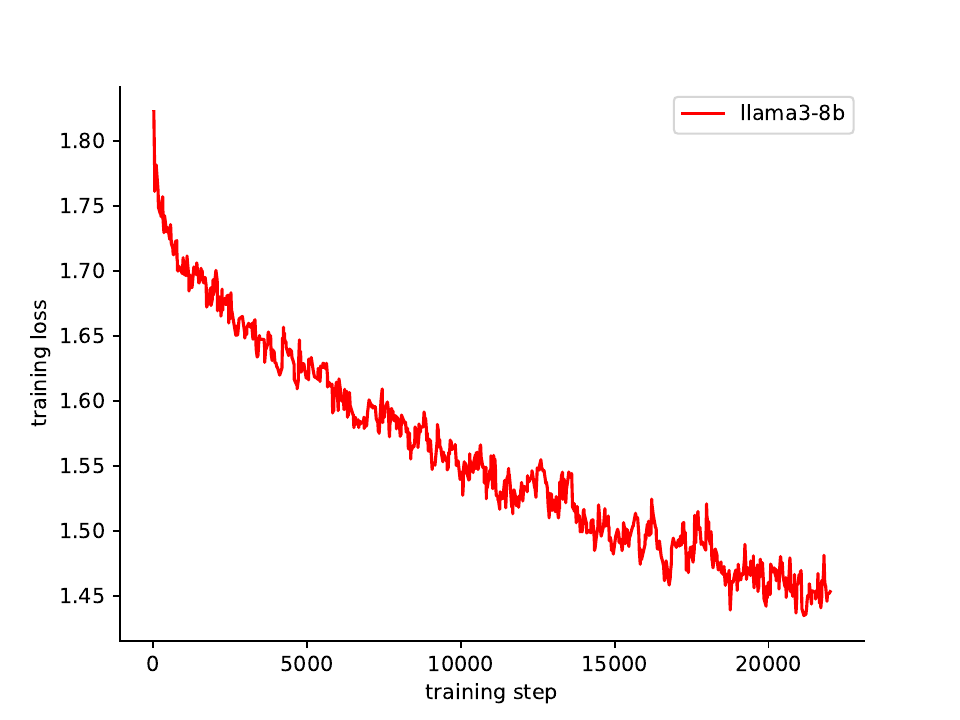}
    \caption{Continue pretraining loss of ChipExpert}
    \label{fig:enter-label}
\end{wrapfigure}

\section{Training Details}

ChipExpert is constructed based on Llama3-8B, utilizing its tokenizer and the standard autoregressive language model objective function for continue pretraining, and the Cross-Entropy loss function for supervised fine-tuning. We selected the ModelLink\footnote{https://gitee.com/ascend/ModelLink} framework for training on 8 Ascend-910B (64GB NPU). We leverage Flash Attention\cite{Dao2022} and Group Query Attention (GQA)\cite{Ainslie2023} from Llama3 to enhance ChipExpert training and inference efficiency. 

\minisection{Continue Pretraining}
During the continue pretraining phase, the learning rate was set to $5 \times 10^{-5}$, an order of magnitude lower than the $10^{-4}$ used in Llama3 pre-training to ensure that the language model learns domain-specific knowledge while avoiding catastrophic forgetting of general knowledge. The global batch size was set to 64, and the number of training steps was set to 25,512 for one epoch on the continue pretraining dataset introduced in §2.1. The whole continue pretraining process was completed in 7 days.

\minisection{Supervised Fine-tuning}
When it comes to the supervised fine-tuning phase, hyperparameters remain the same as in the continue pretraining phase, except for the learning rate, which is set to $5 \times 10^{-6}$. We fine-tune ChipExpert for 2 epochs on a combined dataset of domain-specific and general multi-turn QA pairs after random shuffling. It became apparent that the model rapidly exhibited signs of overfitting when increasing the number of epochs, as evidenced by the model's performance on the test dataset.

\minisection{Two-phase DPO}
For the final phase, we conduct a two-phase Direct Preference Optimization (DPO) process utilizing the human preference alignment data outlined in §2.3. The optimization involves QLoRA\cite{Dettmers2024} fine-tuning, targeting all linear layers with a LoRA scaling factor alpha of 128, a rank of 128, and a dropout rate of $5 \times 10^{-2}$.

\begin{table}[ht]
    \small
    \centering
    \begin{tabular}{l|l|l|l|l}
        \hline
        Learning rate & Optimizer & Seq. len & Global BS & Warmup \\
        \hline
        5.00E-06 & paged\_adamw\_8bit & 1024 & 32 & 10 \\     
        \hline
    \end{tabular}
    \caption{Training details of the DPO stage}
    \label{tab:training_costs}
\end{table}

As demonstrated in Figure 4, the continue pretraining loss of ChipExpert under the specified settings exhibits a consistent downward trend throughout the training process. This continuous decline in loss highlights the efficiency of the training strategy and underscores the model's capability to learn intricate patterns and details from the continue pre-training dataset.

\section{Retrieval-Augmented Generation}

\begin{figure}
    \centering
    \includegraphics[width=0.95\linewidth]{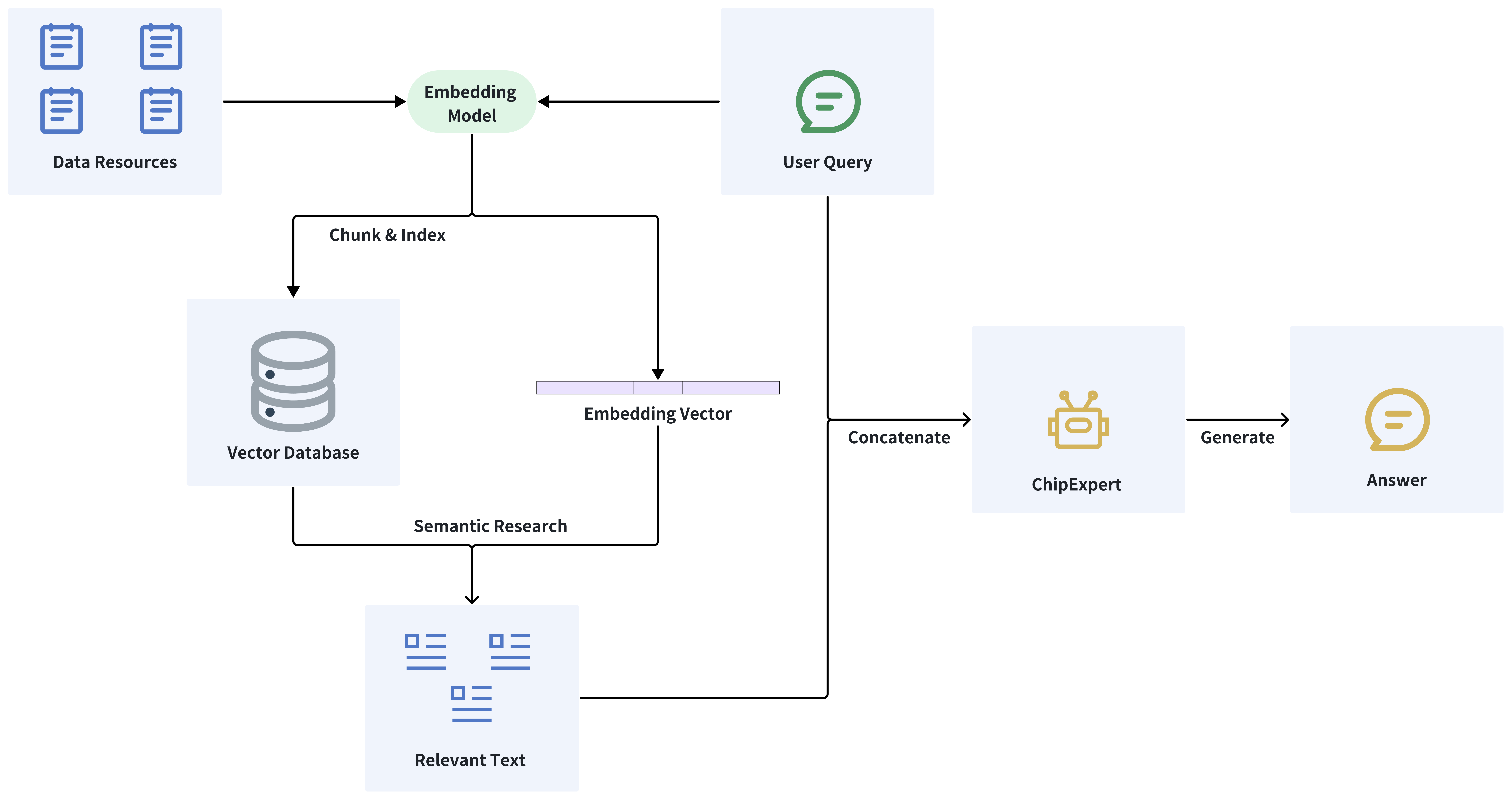}
    \caption{Retrieval-Augmented Generation Workflow}
    \label{fig:human_foundational}
\end{figure}

Retrieval-Augmented Generation (RAG)\cite{{Lewis2020}} can help ChipExpert mitigate hallucinations and obtain the latest and most comprehensive IC design-related information from the knowledge base. As shown in Figure 5, we preprocess domain data from different sources and convert them into embedding vectors using an Embedding Model, which is then stored in a vector database. When a user asks a question, it is first transformed into an embedding vector through the Embedding Model. Next, using Approximate Nearest Neighbor Search (ANN)\cite{Li2019}, the top 3 most similar embedding vectors and their corresponding text passages are retrieved from the vector database. Finally, these text passages are concatenated with the user's question as context and fed into ChipExpert to generate the final answer, which is likely to be better than the one directly generated by ChipExpert.

\section{Evaluations}

In this section, we evaluate the performance of ChipExpert. To assess the effectiveness of our training methodology and the model's application performance, we constructed ChatICD-Bench, a benchmark specifically designed for the IC design domain. We compare the performance of ChipExpert against Llama3-8B and the state-of-the-art language model GPT-4, employing automatic evaluation and human evaluation to ensure that the model’s performance is thoroughly and fairly assessed. Evaluations are conducted using both human and automatic methods. Human evaluation involves experts rating the responses, while automatic evaluation employs a multi-agent scoring model and a referee debate model to provide comprehensive and objective results.

\subsection{ChatICD-Bench: The First IC Design Benchmark}

To quantitatively evaluate the performance of LLMs in IC design tasks, we developed ChatICD-Bench, the first benchmark specifically designed for the IC design field. ChatICD-Bench consists of two types of evaluation questions: foundational questions and advanced questions. To ensure diversity and comprehensive coverage, all subjective questions are meticulously crafted by domain experts.

\begin{figure}
    \centering
    \includegraphics[width=0.75\linewidth]{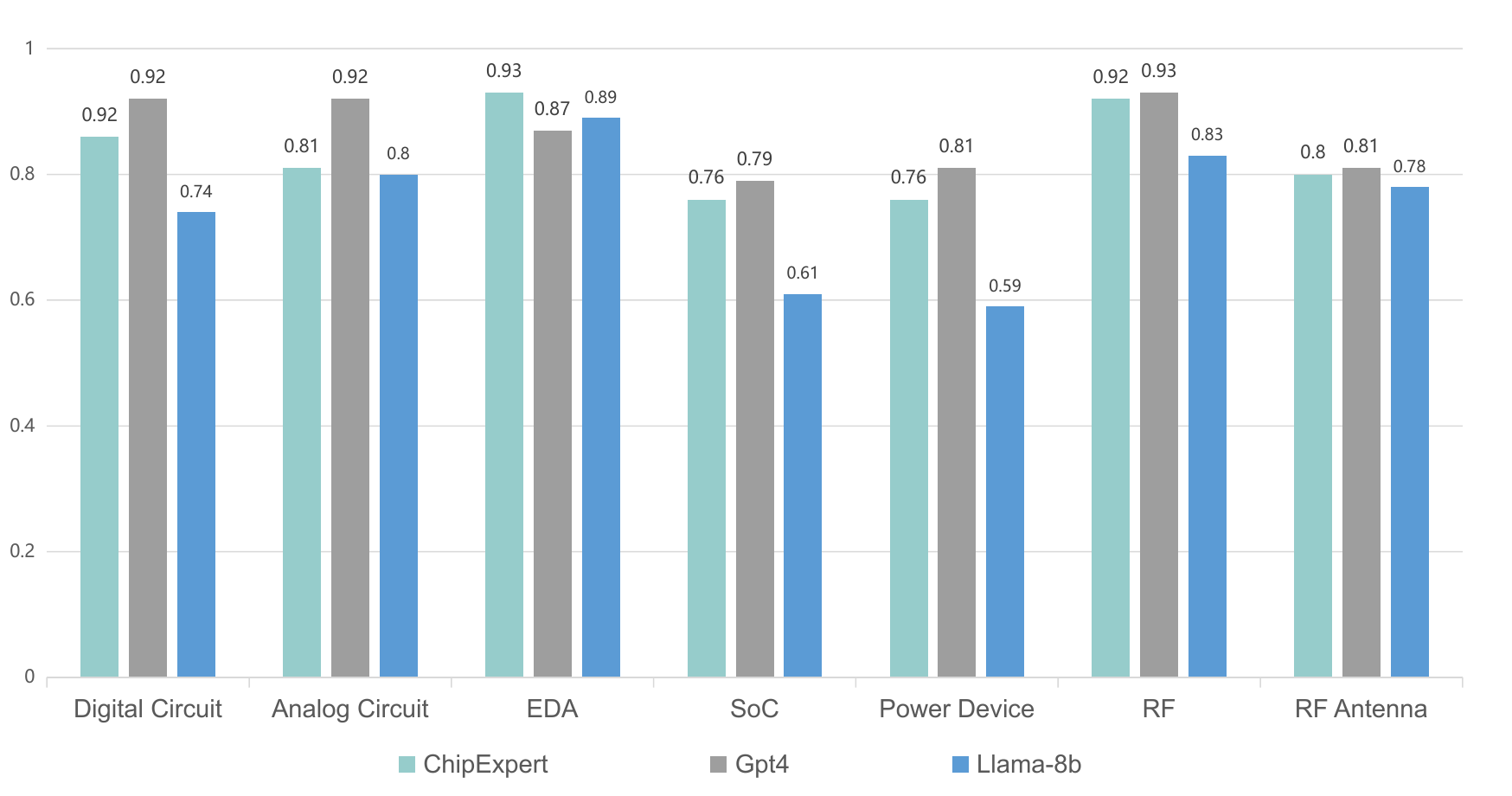}
    \caption{Performance comparison results with human evaluation of ChipExpert, GPT-4, and Llama3-8B on foundational questions from ChatICD-Bench.}
    \label{fig:human_foundational}
\end{figure}

\begin{figure}
    \centering
    \includegraphics[width=0.75\linewidth]{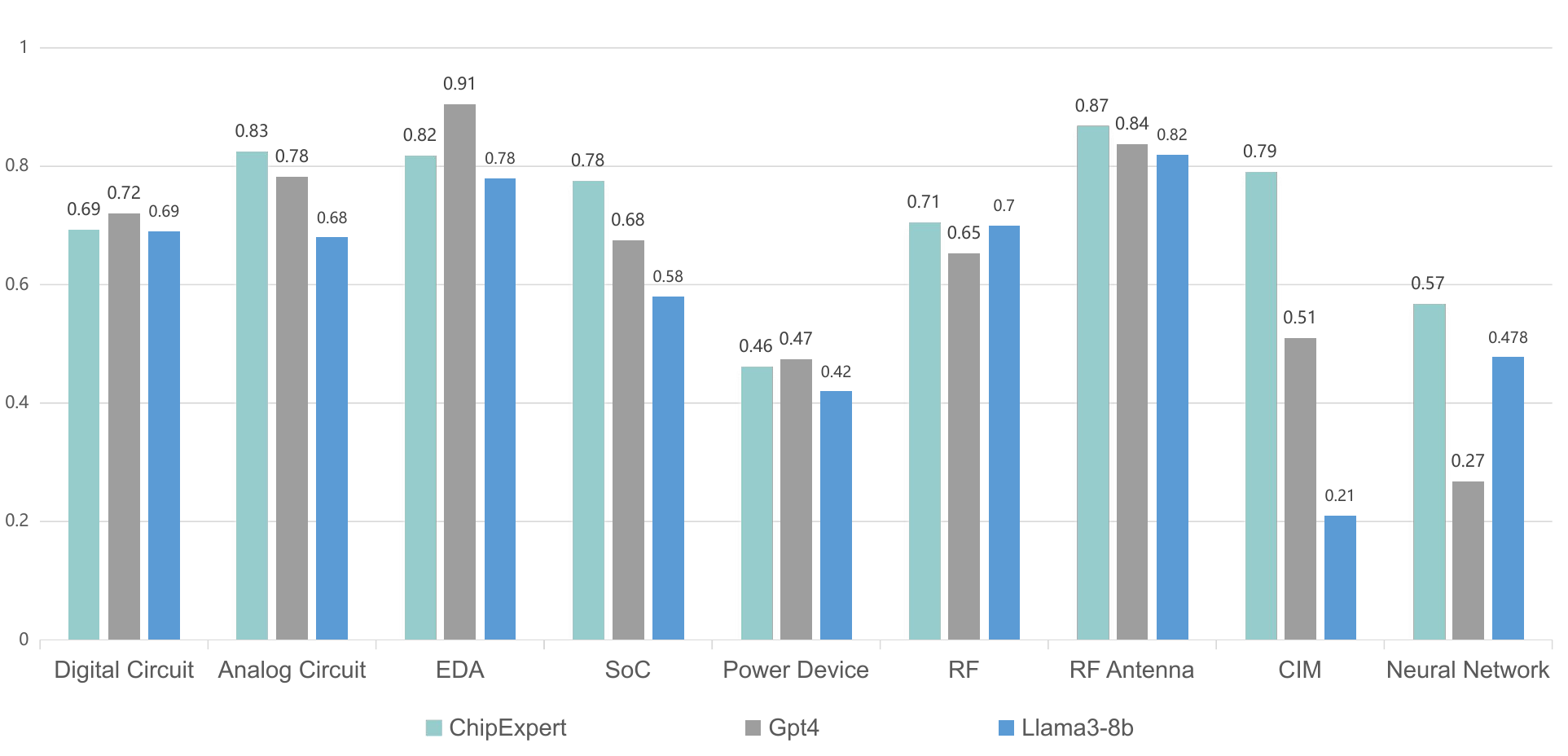}
    \caption{Performance comparison results with human evaluation of ChipExpert, GPT-4, and Llama3-8B on advanced questions from ChatICD-Bench.}
    \label{fig:human_advanced}
\end{figure}

\minisection{Foundational Questions}
The foundational questions in ChatICD-Bench are primarily sourced from classic textbooks. These questions span 7 sub-domains, including analog circuit design, digital circuit design, EDA, SoC, power device design, RF circuit design, and RF antenna design. The foundational questions are designed to evaluate the model's basic capabilities in IC design, covering fundamental concepts and essential problem-solving skills for the field.

\minisection{Advanced Questions}
The advanced questions are derived from recent high-quality research papers. These questions encompass the same sub-domains as the foundational questions, along with two additional sub-domains: compute-in-memory and neural networks. The advanced questions aim to evaluate the model's ability to understand and engage with cutting-edge topics in IC design, testing its proficiency and potential in the latest advancements and innovative approaches within the domain.

\minisection{Human Evaluation}
Human evaluation involves experts rating the responses. Experts in the IC design domain rate the responses on a scale from 0 to 1, considering various criteria such as accuracy, completeness, and relevance. This approach captures subtle aspects of the responses to ensure a thorough and fair assessment.

\minisection{Automatic Evaluation}
To reduce the cost and time associated with human evaluations, we developed an automatic evaluation system that employs an LLM-based approach. This system includes a multi-agent scoring model and a referee debate model. The multi-agent scoring model consists of several agents, each responsible for evaluating different aspects of the response, such as key point redundancy, overall logic, and terminology explanation. The referee debate model assesses the quality of multiple responses through interactive debates to determine the best output. By integrating scores from both the multi-agent scoring model and the referee debate model, we ensure a comprehensive and objective evaluation of the responses while minimizing the reliance on human evaluators.

\subsection{Results and Analysis}
In this section, we present the results and analysis of ChipExpert's performance on ChatICD-Bench compared to Llama3-8B and the state-of-the-art language model GPT-4, utilizing human evaluation methods.

\minisection{Results and Analysis on Foundational Questions}
We compared the performance of ChipExpert with Llama3-8B and GPT-4 on foundational questions from ChatICD-Bench through human evaluation. As shown in Figure \ref{fig:human_foundational}, ChipExpert shows a significant improvement over the Llama3-8B base model in foundational questions. Furthermore, ChipExpert with 8B parameters achieves a comparable performance to GPT-4 in the IC design domain. Notably, in foundational questions within the EDA domain, ChipExpert scored 0.93, surpassing GPT-4's score of 0.87. This result demonstrates that ChipExpert outperforms the state-of-the-art GPT-4 in foundational EDA questions, achieving a high level of proficiency in the EDA domain. In other sub-domains, except for Analog Circuit, ChipExpert's performance was close to GPT-4, indicating that ChipExpert is comparable to GPT-4 for most foundational questions across various IC design sub-domains. The performance gap in the analog circuit design domain can be attributed to the complexity and specialized nature of analog circuit design, which may not have been as thoroughly represented in the pre-training data. 

\minisection{Results and Analysis on Advanced Questions}
Additionally, we compared the performance of ChipExpert with Llama3-8B and GPT-4 on advanced questions from ChatICD-Bench through human evaluation. As illustrated in Figure \ref{fig:human_advanced}, ChipExpert demonstrates superior performance in all advanced IC design domains compared to Llama3-8B. It is notable that ChipExpert outperformed GPT-4 in 6 out of 9 advanced sub-domains. Particularly, in emerging fields such as compute-in-memory, ChipExpert outperforms that of GPT-4 by 0.28, highlighting its capability to excel in cutting-edge areas within the IC design domain. The superior performance of ChipExpert compared to GPT-4 in the most advanced sub-domains can be attributed to the continued pretraining data with an extensive collection of recent high-quality research papers.

\section{Related Works}
\minisection{Domain-Specific Language Model Advancements}
The integration of LLMs into specialized domains has been a significant development in the field of artificial intelligence. The potential of LLMs to revolutionize various industries is evident through the emergence of domain-specific models that cater to the unique needs of their respective fields. In the financial sector, BloombergGPT\cite{Wu2023} has emerged as a pioneering model, showcasing how LLMs can be fine-tuned to understand and predict complex financial trends. Similarly, OceanGPT\cite{Bi2023} has been tailored to address the linguistic and knowledge intricacies of ocean science, demonstrating the adaptability of LLMs to specialized domains. The medical field has also seen the introduction of PMC-LLaMA\cite{Wu2024}, an open-source LLM designed to assist in medical research and clinical decision-making, underlining the growing trend of LLMs becoming indispensable tools in professional domains. These models have not only enhanced the efficiency of information processing but also raised the bar for the level of domain-specific knowledge that LLMs must possess.

\minisection{Technological Frameworks for LLMs}
The technical underpinnings of LLMs have been bolstered by innovations in training methodologies and attention mechanisms. The work on continuing pretraining by Gururangan\cite{Gururangan2020} has set a precedent for models like ChipExpert, which continue to learn and adapt to new information while retaining foundational knowledge. The introduction of techniques such as flash attention\cite{Dao2022} and GQA\cite{Ainslie2023} has improved the training efficiency of LLMs, allowing them to process vast amounts of data more effectively.RAG\cite{Lewis2020} has been a game-changer in addressing the limitations of LLMs, such as hallucination, by enabling them to retrieve and integrate accurate information from a knowledge base. Ethical considerations have also come to the forefront with DPO\cite{Rafailov2024}, ensuring that LLMs align with human preferences and societal norms, thus mitigating potential risks. The scaling laws\cite{Kaplan2020} and studies on catastrophic forgetting\cite{Lin2023} provide critical insights into the balance between general and domain-specific knowledge, guiding the development of models that can adapt to new domains without losing their foundational understanding.

\section{Conclusion}
In this paper, we introduced ChipExpert, the first open-source instructional large language model specifically tailored for the IC design domain. Our comprehensive training process included continued pretraining on IC-specific data and supervised fine-tuning with domain-specific question-answer pairs. We developed ChatICD-Bench, the first benchmark designed to evaluate the performance of LLMs in IC design tasks, and used it to compare ChipExpert with GPT-4. Our evaluation results demonstrate that ChipExpert, with 8B parameters, achieves a comparable performance to GPT-4. Our future work will focus on training ChipExpert on larger base models and exploring multimodal large language model(MLLM) in the IC design domain to further enhance its capabilities and performance across all facets of IC design.

\appendix
\section{Appendix}
\subsection{Examples pf ChipExpert's Response to Foundational Questions}
\begin{tcolorbox}[width=1.0\linewidth, halign=left, colframe=black, colback=white, boxsep=0.01mm, arc=1.5mm, left=2mm, right=2mm, boxrule=0.5pt, breakable]\footnotesize
\textbf{Question:}\\
Differences between CST and HFSS and the advantages and disadvantages.

\vspace{0.2in}
\textbf{ChipExpert Answer:}\\
CST Microwave Studio (CST) and High Frequency Structure Simulator (HFSS) are both widely used electromagnetic simulation software tools for modeling and simulating high-frequency structures and devices. While they share many similarities in their capabilities and features, there are some key differences between them:

1. User Interface: The user interface of CST is more intuitive and easier to navigate compared to HFSS. CST has a more modern look and feel, while HFSS's interface may seem more cluttered and complex.

2. Meshing: CST uses a hybrid meshing approach, which combines tetrahedral and hexahedral elements. This allows it to handle complex geometries better than HFSS, which primarily uses tetrahedral elements. However, this also means that CST requires more computational resources and time to generate the mesh.

3. Material Modeling: Both CST and HFSS support a wide range of material models, including frequency-dependent and dispersive materials. However, CST has a more comprehensive set of material libraries and supports more advanced material models.

4. Post-Processing: CST offers a more robust post-processing capability, allowing users to analyze and visualize results in greater detail. It also includes built-in tools for optimizing designs and generating reports.

5. Support: CST has a larger global user base and a more active community, resulting in faster response times when seeking technical support or assistance from the vendor.

In summary, both CST and HFSS have their strengths and weaknesses, and the choice between them often depends on factors such as the specific application, user experience, and available computational resources.
\end{tcolorbox}

\begin{tcolorbox}[width=1.0\linewidth, halign=left, colframe=black, colback=white, boxsep=0.01mm, arc=1.5mm, left=2mm, right=2mm, boxrule=0.5pt, breakable]\footnotesize
\textbf{Question:}\\
When reverse engineering a power chip, after knowing its general structure, it is necessary to analyze the doping concentration and distribution in the drift region of the device. What are the common methods used for this analysis?

\vspace{0.2in}
\textbf{ChipExpert Answer:}\\
To accurately determine the doping concentration and distribution in the drift region of a power chip, several techniques can be employed:

1. Spreading resistance profiling (SRP): This method involves measuring the resistance between two contacts on opposite sides of the chip while gradually increasing the spacing between them. The resulting data is then processed using mathematical models to extract the doping profile.

2. Capacitance-voltage (C-V) measurements: By applying a voltage across the device and measuring the corresponding capacitance, the doping profile can be inferred from the relationship between depletion width and applied voltage.

3. Secondary ion mass spectrometry (SIMS): In this technique, ions are sputtered from the surface of the chip at a specific rate, and the resulting secondary ions are analyzed using a mass spectrometer. The depth resolution depends on the sputtering rate, allowing for precise measurement of the doping profile.

4. Scanning spreading resistance microscopy (SSRM): Similar to SRP, SSRM uses a sharp tip to measure the resistance between two points on the chip surface. However, instead of varying the distance between the contacts, the tip position is scanned across the surface, providing a detailed map of the local resistance and, consequently, the doping profile.

These methods enable researchers to characterize the critical drift region of power chips, which is essential for understanding their electrical properties and optimizing their performance.
\end{tcolorbox}

\begin{tcolorbox}[width=1.0\linewidth, halign=left, colframe=black, colback=white, boxsep=0.01mm, arc=1.5mm, left=2mm, right=2mm, boxrule=0.5pt, breakable]\footnotesize
\textbf{Question:}\\
How can machine learning and artificial intelligence techniques be leveraged to improve the accuracy and efficiency of power estimation and optimization in EDA tools, considering the growing importance of power management in modern electronic systems?

\vspace{0.2in}
\textbf{ChipExpert Answer:}\\
Machine learning (ML) and artificial intelligence (AI) techniques have emerged as powerful tools for enhancing the accuracy and efficiency of power estimation and optimization in electronic design automation (EDA) tools. As power consumption has become a critical concern in modern electronic systems, particularly in battery-powered devices and high-performance computing applications, the ability to accurately estimate and optimize power usage has gained paramount importance.

One key application of ML/AI in power estimation is the development of predictive models that leverage historical data and trends to forecast power consumption more precisely than traditional analytical methods. These models can learn from past designs, fabrication processes, and environmental factors to capture complex relationships between circuit parameters and power dissipation. By incorporating advanced algorithms such as neural networks, decision trees, or support vector machines, these models can provide highly accurate predictions, even in the presence of noise or uncertainties in the input data.

Moreover, AI techniques can be employed to automate the process of identifying and optimizing low-power design strategies. By analyzing vast amounts of design data and exploring various design alternatives, AI algorithms can discover novel approaches to reducing power consumption while maintaining performance and functionality. This can involve techniques like voltage scaling, clock gating, power gating, or the use of energy-efficient architectures. The AI system can iteratively refine its recommendations based on feedback from simulations or actual hardware measurements, continuously improving the effectiveness of the power optimization process.

In addition to power estimation and optimization, ML/AI can also play a crucial role in accelerating the design exploration phase. By rapidly evaluating numerous design configurations and trade-offs, AI-driven tools can help designers navigate the complex landscape of power-performance-area (PPA) metrics. This enables faster convergence towards optimal solutions, reducing the time-to-market and improving overall design productivity.

Furthermore, ML/AI techniques can facilitate the integration of power-awareness into higher-level design flows, such as system-on-chip (SoC) design or software-hardware co-design. By providing intelligent guidance and decision-making capabilities, AI systems can assist designers in making informed choices regarding power management at various levels of abstraction, from architecture to implementation.

The successful deployment of ML/AI in EDA tools requires careful consideration of several aspects. Firstly, the quality and quantity of training data are essential for building robust and reliable models. Secondly, the choice of appropriate algorithms and model architectures must align with the specific requirements of power estimation and optimization tasks. Thirdly, interpretability and explainability of the AI decisions are crucial for gaining designer trust and enabling effective collaboration between humans and machines.

By leveraging the capabilities of ML/AI, EDA tools can significantly enhance their power estimation and optimization capabilities, leading to more efficient and sustainable electronic systems. This will enable designers to create innovative products that meet the ever-increasing demands for low power consumption, extended battery life, and thermal management. As the field of EDA continues to evolve, the integration of ML/AI techniques will undoubtedly play a pivotal role in shaping the future of power-aware design practices.
\end{tcolorbox}

\subsection{Examples pf ChipExpert's Response to Advanced Questions}

\begin{tcolorbox}[width=1.0\linewidth, halign=left, colframe=black, colback=white, boxsep=0.01mm, arc=1.5mm, left=2mm, right=2mm, boxrule=0.5pt, breakable]\footnotesize
\textbf{Question:}\\
How to unify knowledge distillation (KD) and self-knowledge distillation (self-KD) using NKD (Normalized KD) and USKD (Universal Self-KD) without relying on a real teacher model to generate soft labels

\vspace{0.2in}
\textbf{ChipExpert Answer:}\\
What a fascinating topic!

Knowledge Distillation (KD) is a technique that transfers knowledge from a pre-trained teacher model to a student model by minimizing the difference between their outputs. Self-Knowledge Distillation (Self-KD), on the other hand, involves training a single model to mimic its own predictions. Normalized Knowledge Distillation (NKD) and Universal Self-KD (USKD) are two variants of these techniques.

To unify KD and Self-KD without relying on a real teacher model, you can use the following approach:

1. \textbf{Normalize the student model's output}: Normalize the output of the student model to ensure it produces probability distributions over the same space as the target distribution (e.g., uniform or categorical). This helps in aligning the student's predictions with the target distribution.

2. \textbf{Use a proxy task}: Design a proxy task that mimics the behavior of the teacher model. For example, if the teacher model is a classification model, the proxy task could be a regression problem where the goal is to predict the class probabilities.

3. \textbf{Train the student model on the proxy task}: Train the student model on the proxy task using the normalized output as the target. This will help the student model learn to produce similar outputs as the teacher model.

4. \textbf{Apply USKD}: Use the trained student model as the teacher model for the original task. This is known as Universal Self-KD (USKD). The student model will now learn to mimic its own predictions, which were previously learned through the proxy task.

5. \textbf{Fine-tune the student model}: Fine-tune the student model on the original task using the USKD loss function. This will help refine the student model's performance on the original task.

Here's a mathematical formulation of the process:

Let $p(y|x)$ be the true probability distribution over the output space, $q_\theta(x)$ be the student model's output, and $r(x)$ be the normalized output of the student model.

\textbf{Proxy Task Loss}:
$$L_{proxy} = \mathbb{E}_{x\sim D}[\ell(r(x), p(y|x))]$$

where $\ell$ is a loss function (e.g., cross-entropy).

\textbf{USKD Loss}:
$$L_{uskd} = \mathbb{E}_{x\sim D}[\ell(q_\theta(x), r(x))]$$

The overall loss function combines both losses:
$$L = L_{proxy} + \alpha L_{uskd}$$

where $\alpha$ is a hyperparameter controlling the trade-off between the two losses.

By unifying KD and Self-KD using NKD and USKD, you can leverage the strengths of both approaches without requiring a real teacher model. This method has been shown to improve the performance of student models in various tasks, such as image classification and language modeling.

I hope this explanation helps! Do you have any further questions or would you like me to elaborate on any specific aspect?
\end{tcolorbox}

\begin{tcolorbox}[width=1.0\linewidth, halign=left, colframe=black, colback=white, boxsep=0.01mm, arc=1.5mm, left=2mm, right=2mm, boxrule=0.5pt, breakable]\footnotesize
\textbf{Question:}\\
What are the main challenges in implementing XOR-gate compression for transformer models, and how can these challenges be addressed to achieve stall-free sparse-model access?

\vspace{0.2in}
\textbf{ChipExpert Answer:}\\
XOR-gate compression is a technique used to compress deep neural networks by replacing dense weights with sparse binary weights using XOR gates. Implementing XOR-gate compression for transformer models poses several challenges, which I'll outline below:

1. \textbf{Computational overhead}: XOR gates introduce additional computational complexity due to the need to compute the bitwise XOR operation at each layer. This can lead to slower inference times.

2. \textbf{Memory access patterns}: Sparse binary weights require non-contiguous memory access patterns, which can result in cache misses and decreased performance.

3. \textbf{Training difficulties}: Training XOR-gate compressed models can be challenging due to the discrete nature of the binary weights, making it harder to optimize the model during training.

4. \textbf{Quantization errors}: Binary weights can lead to quantization errors, especially when dealing with floating-point operations.

To address these challenges and achieve stall-free sparse-model access, consider the following strategies:

1. \textbf{Optimize hardware architecture}: Design specialized hardware accelerators or modify existing ones to efficiently handle XOR gate computations and sparse memory access patterns.

2. \textbf{Cache-friendly data structures}: Use data structures that minimize cache misses, such as block-based storage or hierarchical caching, to reduce memory access latency.

3. \textbf{Efficient training algorithms}: Develop training algorithms that take into account the discrete nature of binary weights, such as stochastic gradient descent (SGD) with adaptive learning rate or binary weight pruning.

4. \textbf{Error compensation techniques}: Implement error compensation techniques, like quantization-aware training or learned quantization, to mitigate the effects of quantization errors.

5. \textbf{Model pruning and knowledge distillation}: Apply model pruning techniques to remove redundant connections and knowledge distillation to transfer knowledge from the original dense model to the compressed one.

6. \textbf{Hybrid approaches}: Combine XOR-gate compression with other compression techniques, such as pruning or knowledge distillation, to further improve compression ratios and accuracy.

7. \textbf{Software optimizations}: Optimize software frameworks and libraries to efficiently support XOR-gate compressed models, including optimized kernels for matrix multiplication and convolutional operations.

8. \textbf{Hardware-software co-design}: Collaborate on hardware-software co-design efforts to develop custom accelerators that seamlessly integrate with software frameworks and optimize for XOR-gate compressed models.

By addressing these challenges and incorporating these strategies, you can effectively implement XOR-gate compression for transformer models and achieve stall-free sparse-model access.
\end{tcolorbox}

\begin{tcolorbox}[width=1.0\linewidth, halign=left, colframe=black, colback=white, boxsep=0.01mm, arc=1.5mm, left=2mm, right=2mm, boxrule=0.5pt, breakable]\footnotesize
\textbf{Question:}\\
What are the benefits of hybrid CIM? Why do researchers want to combine the analog and digital methods together?

\vspace{0.2in}
\textbf{ChipExpert Answer:}\\
Hybrid Computing-in-Memory (CIM) architectures represent a significant advancement in computing technology, combining the strengths of both analog and digital methods within a single system. This innovative approach aims to address some of the key challenges faced by traditional computing systems, such as power consumption, latency, and scalability.

In conventional computing systems, data processing typically involves moving information between separate memory units and processing units. However, this process can be time-consuming and energy-intensive, especially when dealing with large datasets or complex computations. By integrating computation directly into the memory itself, CIM architectures aim to minimize these inefficiencies by reducing the need for data movement.

The analog component of hybrid CIM leverages the physical properties of memory cells to perform certain types of operations more efficiently than their digital counterparts. For example, analog circuits can perform matrix-vector multiplications, which form the backbone of many machine learning algorithms, with high speed and low power consumption. This makes them particularly well-suited for applications like neural network inference, where rapid processing of large amounts of data is crucial.

On the other hand, the digital component of hybrid CIM ensures the reliability and flexibility required for more complex tasks. Digital circuits excel at precise arithmetic operations, error correction, and control functions, providing the robustness necessary for tasks like training deep learning models or running complex software.

By combining these two approaches, hybrid CIM architectures offer a powerful solution that can adapt to the specific requirements of different workloads. The analog components can handle the bulk of the data processing, while the digital components provide the necessary safeguards and fine-grained control. This synergistic combination allows for faster, more efficient processing of data, making it possible to tackle increasingly complex computational problems with greater ease.
\end{tcolorbox}

\begin{tcolorbox}[width=1.0\linewidth, halign=left, colframe=black, colback=white, boxsep=0.01mm, arc=1.5mm, left=2mm, right=2mm, boxrule=0.5pt, breakable]\footnotesize
\textbf{Question:}\\
Describe the work principle of the conventional buck-boost converter.

\vspace{0.2in}
\textbf{ChipExpert Answer:}\\
The conventional buck-boost converter is a type of DC-DC converter that can step up or step down the input voltage while maintaining a constant output current. It's widely used in various applications such as power supplies, motor drives, and renewable energy systems.

The basic working principle of a buck-boost converter can be explained as follows:

\textbf{Components:}

1. Input capacitor (Cin)
2. Switching transistor (Q1)
3. Diode (D1)
4. Output capacitor (Cout)
5. Inductor (L)

\textbf{Operation:}

The converter operates in two modes: Buck mode and Boost mode.

\textbf{Buck Mode:}
When the switching transistor Q1 is turned on, it connects the input voltage Vin to the inductor L, causing the current to increase. The diode D1 is reverse-biased, and the output capacitor Cout is charged through the inductor L. This process is known as "inductive charging."

As the current in the inductor increases, the voltage across the inductor also increases, which causes the output voltage Vout to decrease. When the switching transistor Q1 is turned off, the inductor current continues to flow through the diode D1, and the output capacitor Cout discharges. This process is known as "inductive discharge."

\textbf{Boost Mode:}
When the switching transistor Q1 is turned off, it disconnects the input voltage Vin from the inductor L, allowing the inductor current to collapse. The diode D1 becomes forward-biased, and the output capacitor Cout is discharged through the inductor L. This process is known as "inductive discharge."

As the current in the inductor collapses, the voltage across the inductor also decreases, which causes the output voltage Vout to increase. When the switching transistor Q1 is turned on again, the inductor current starts to increase, and the process repeats.

\textbf{Key Benefits:}

1. \textbf{Voltage regulation:} The buck-boost converter can regulate the output voltage by adjusting the duty cycle of the switching transistor.

2. \textbf{Efficiency:} The converter has high efficiency due to the use of an inductor, which helps to reduce losses.

3. \textbf{Flexibility:} The buck-boost converter can operate in both buck and boost modes, making it suitable for a wide range of applications.

I hope this explanation helps you understand the working principle of the conventional buck-boost converter! Do you have any further questions?
\end{tcolorbox}

\begin{tcolorbox}[width=1.0\linewidth, halign=left, colframe=black, colback=white, boxsep=0.01mm, arc=1.5mm, left=2mm, right=2mm, boxrule=0.5pt, breakable]\footnotesize
\textbf{Question:}\\
What is the role of a software-controlled stall rate (SSR) in the SOC?

\vspace{0.2in}
\textbf{ChipExpert Answer:}\\
I'd be happy to help you understand the role of Software-Controlled Stall Rate (SSR) in a System-on-Chip (SoC).

In a SoC, the SSR is a critical component that plays a crucial role in managing the power consumption and performance of the system. Specifically, it helps regulate the frequency at which the processor stalls or waits for external events, such as memory access or I/O operations.

When a processor encounters a stall, it means that it needs to wait for some event to complete before continuing execution. This can happen due to various reasons like cache misses, page faults, or waiting for data from peripherals. The SSR is responsible for controlling the rate at which these stalls occur.

The primary function of the SSR is to dynamically adjust the stall rate based on the system's workload, temperature, and voltage/frequency scaling (DVFS). By doing so, it helps:

1. \textbf{Optimize power consumption}: By reducing the number of stalls, the SSR minimizes the time spent in idle states, which reduces power consumption.

2. \textbf{Improve performance}: By allowing the processor to execute more instructions between stalls, the SSR improves overall system performance.

3. \textbf{Enhance thermal management}: By regulating the stall rate, the SSR helps maintain a stable operating temperature, preventing overheating and reducing the risk of throttling.

To achieve this, the SSR typically uses various techniques, including:

* Dynamic voltage and frequency scaling (DVFS)

* Power gating

* Clock gating

* Cache hierarchy management

* Branch prediction and speculation

By carefully controlling the stall rate, the SSR enables the SoC to adapt to changing workloads, ensuring optimal performance, power efficiency, and thermal management.

Would you like me to elaborate on any specific aspect of SSR or its implementation?
\end{tcolorbox}


\begin{thebibliography}{1}

\bibitem{Achiam2023}
Achiam J, Adler S, Agarwal S, et al. Gpt-4 technical report[J]. arXiv preprint arXiv:2303.08774, 2023.

\bibitem{Touv2023}
Touvron H, Lavril T, Izacard G, et al. Llama: Open and efficient foundation language models[J]. arxiv preprint arxiv:2302.13971, 2023.

\bibitem{Touvron2023}
Touvron H, Martin L, Stone K, et al. Llama 2: Open foundation and fine-tuned chat models[J]. arXiv preprint arXiv:2307.09288, 2023.

\bibitem{Gururangan2020}
Gururangan S, Marasović A, Swayamdipta S, et al. Don't stop pretraining: Adapt language models to domains and tasks[J]. arXiv preprint arXiv:2004.10964, 2020.

\bibitem{Rafailov2024}
Rafailov R, Sharma A, Mitchell E, et al. Direct preference optimization: Your language model is secretly a reward model[J]. Advances in Neural Information Processing Systems, 2024, 36.

\bibitem{Lewis2020}
Lewis P, Perez E, Piktus A, et al. Retrieval-augmented generation for knowledge-intensive nlp tasks[J]. Advances in Neural Information Processing Systems, 2020, 33: 9459-9474.

\bibitem{Wu2023}
Wu S, Irsoy O, Lu S, et al. Bloomberggpt: A large language model for finance[J]. arXiv preprint arXiv:2303.17564, 2023.

\bibitem{Bi2023}
Bi Z, Zhang N, Xue Y, et al. Oceangpt: A large language model for ocean science tasks[J]. arXiv preprint arXiv:2310.02031, 2023.

\bibitem{Wu2024}
Wu C, Lin W, Zhang X, et al. PMC-LLaMA: toward building open-source language models for medicine[J]. Journal of the American Medical Informatics Association, 2024: ocae045.

\bibitem{Kaplan2020}
Kaplan J, McCandlish S, Henighan T, et al. Scaling laws for neural language models[J]. arXiv preprint arXiv:2001.08361, 2020.

\bibitem{Kaufmann2023}
Kaufmann T, Weng P, Bengs V, et al. A survey of reinforcement learning from human feedback[J]. arXiv preprint arXiv:2312.14925, 2023.

\bibitem{Rao2016}
Rao B C, Zhu E. Searching web data using minhash LSH[C]//Proceedings of the 2016 International Conference on Management of Data. 2016: 2257-2258.

\bibitem{Dong2023}
Dong G, Yuan H, Lu K, et al. How abilities in large language models are affected by supervised fine-tuning data composition[J]. arXiv preprint arXiv:2310.05492, 2023.

\bibitem{Lin2023}
Lin Y, Tan L, Lin H, et al. Speciality vs generality: An empirical study on catastrophic forgetting in fine-tuning foundation models[J]. arXiv preprint arXiv:2309.06256, 2023.

\bibitem{Ganguli2022}
Ganguli D, Lovitt L, Kernion J, et al. Red teaming language models to reduce harms: Methods, scaling behaviors, and lessons learned[J]. arXiv preprint arXiv:2209.07858, 2022.

\bibitem{Inan2023}
Inan H, Upasani K, Chi J, et al. Llama guard: Llm-based input-output safeguard for human-ai conversations[J]. arXiv preprint arXiv:2312.06674, 2023.

\bibitem{Dao2022}
Dao T, Fu D Y, Ermon S, et al. Fast and memory-efficient exact attention with io-awareness, 2022[J]. URL https://arxiv. org/abs/2205.14135.

\bibitem{Ainslie2023}
Ainslie J, Lee-Thorp J, de Jong M, et al. GQA: Training Generalized Multi-Query Transformer Models from Multi-Head Checkpoints[C]//Proceedings of the 2023 Conference on Empirical Methods in Natural Language Processing. 2023: 4895-4901.

\bibitem{Dettmers2024}
Dettmers T, Pagnoni A, Holtzman A, et al. Qlora: Efficient finetuning of quantized llms[J]. Advances in Neural Information Processing Systems, 2024, 36.

\bibitem{Li2019}
Li W, Zhang Y, Sun Y, et al. Approximate nearest neighbor search on high dimensional data—experiments, analyses, and improvement[J]. IEEE Transactions on Knowledge and Data Engineering, 2019, 32(8): 1475-1488.




\end{thebibliography}
\end{document}